# Fibre Fabry-Pérot Astrophotonic Correlation Spectroscopy for Remote Gas Identification and Radial Velocity Measurements

ROSS CHERITON[1], ADAM DENSMORE[2], SURESH SIVANANDAM[3,4], ERNST DE MOOIJ[5], PAVEL CHEBEN[1], DAN-XIA XU[1], JENS H. SCHMID[1], AND SIEGFRIED JANZ[1]

[1]Advanced Electronics and Photonics, National Research Council Canada, Ottawa, Canada
[2]Herzberg Astronomy and Astrophysics, National Research Council Canada, Victoria, Canada
[3]David A. Dunlap Department of Astronomy and Astrophysics, University of Toronto, Toronto, Canada
[4]Dunlap Institute for Astronomy and Astrophysics, University of Toronto, Toronto, Canada
[5]Astrophysics Research Center, School of Mathematics and Physics, Queen's University Belfast, Belfast, UK

*ross.cheriton@nrc-cnrc.gc.ca

**Abstract:** We present a novel remote gas detection and identification technique based on correlation spectroscopy with a piezoelectric tunable fibre-optic Fabry-Pérot filter. We show that the spectral correlation amplitude between the filter transmission window and gas absorption features is related to the gas absorption optical depth, and that different gases can be distinguished from one another using their correlation signal phase. Using an observed telluric-corrected, high-resolution near-infrared spectrum of Venus, we show via simulation that the Doppler shift of gases lines can be extracted from the phase of the lock-in signal using low-cost, compact, and lightweight fibre-optic components with lock-in amplification to improve the signal-to-noise ratio. This correlation spectroscopy technique has applications in the detection and radial velocity determination of faint spectral features in astronomy and remote sensing. We experimentally demonstrate remote $CO_2$ detection system using a lock-in amplifier, fibre-optic Fabry-Pérot filter, and single channel photodiode.

## 1. Introduction

Passive remote gas sensing is required where laser-based remote sensing either becomes impractical or infeasible to probe distant targets. For astronomical targets, atomic and molecular composition is characterized through absorption [1]or emission spectroscopy [2], where a distribution of spectral lines represent unique "fingerprints" of their molecular or atomic constituents. Dispersive spectroscopy is frequently used to separate the incoming light by wavelength onto a multichannel detector to measure the spectrum, which is then processed and interpreted to infer the gas target composition. Such methods are limited to bright sources or long acquisition times to obtain a sufficiently high signal-to-noise ratio (SNR) to confirm detection, and suffer from high costs and large instrument sizes. To determine faint spectral features in remote sensing, the measured spectrum can be cross-correlated with a reference spectrum to search for a significant signal at the point of optimum overlap [3–5]. Cross-correlation spectroscopy is typically carried out in a numerical post-processing step using model generated reference spectra, but the data acquisition still suffers from the low throughput and SNR ratios introduced by dispersive spectrometers. This technique is used frequently for remote gas sensing in astronomy due to the typical low SNR for many targets.

Cavity-based spectroscopic techniques, such as differential optical absorption spectroscopy [6,7] can improve sensitivity using a comparison between the measured data with a reference spectrum, but spectral acquisition on a linear detection array is still required. Dual-comb spectroscopy is a correlation-based technique that uses two frequency combs to produce a radio-frequency signal for heterodyne detection, but suffers from high cost, limited range, and significant electrical complexity [8–11]. Fabry-Pérot (FP) filters are a type of resonant cavity that generate longitudinal transmission modes at resonant wavelengths [12]. FP interferometers have been used for spectroscopy of remote gases by modulating the cavity length and observing the modulated signal [13–15]. Recently, we have alternatively demonstrated correlation spectroscopy based remote photonic sensing using a silicon waveguide ring resonator on a silicon-on-insulator integrated photonic chip [16].

While FP cavities have existed for many years in bulk optical systems, recent implementations have been realized on fibre-optic platforms. Fibre optic-based technologies hold advantages over their bulk equivalents in regard to routing, cost, weight, propagation distance, size, and stability. Fibre FP cavities have been used in a wide variety of optical sensors, including for temperature, local gas [17], white light interferometry [18], pressure [19], strain [20], ultrasound [21,22], magnetic field [20], flow [23], refractive index [17], vibration [24], and voltage [25]. Tunable fibre FP (FFP) cavities have been recently realized through mechanical tuning of the gap between two cleaved fibre ends. The separation between each fibre facet can be controlled piezo-electrically [25], thermally [26], through microelectromechanical systems (MEMS) [27], or left uncontrolled such that the transmission wavelength is the measurement parameter [28]. FFPs have been realized for local detection of gas in hollow core fibres [29], but have not to our knowledge been demonstrated for remote gas sensing.

Molecular absorption features span the visible, near infrared and mid-infrared spectrum. Sensing of remote gases involves determining spectral features to infer the gas presence of a gas. Dispersive spectroscopy is routinely used to measure the spectrum to directly observe the spectral features. However, the SNR in dispersive spectroscopy is diminished through the division of photons by wavelength into a detection array. Also, losses from the use of diffraction gratings and other optical elements limit the typical throughput of high-resolution spectrographs to ~10-20%. This throughput limitation is encountered when molecular spectroscopy is desired as the ability to detect and distinguish absorption or emission lines requires high spectral resolution. In contrast, photometric techniques which combine all light in spectral bandwidth onto a single detection element, are usually inherently simpler, leading to generally higher throughput, and can operate with a lock-in amplifier with extremely low SNR conditions. The lock-in amplifier can reduce the noise bandwidth significantly by rejecting signals from outside the modulation frequency at the expense of dynamic response, a desirable trade-off in relatively static gas targets. Since the gas composition in Earth's atmosphere is relatively stable over short time periods, the quantification of trace gases such as $CO_2$, $CH_4$, $CO$ and $NO_2$ is a suitable application for this detection method.

Many state-of-the-art astronomical telescopes now also host adaptive optics systems capable of efficient coupling of light into single mode fibres, enabling the use of astrophotonic instruments such as FFP filters. An application in astronomy could involve exoplanet atmosphere characterization, where detecting weak signals are to be detected in strong starlight background. In this case, absorption line depths can be $<10^{-4}$ relative to the star background light, with Earth-like exoplanet detection requiring absorption line depths of $<10^{-6}$ [30–32].. The use of a lock-in amplifier is well-suited to extract the weak modulated signals from the correlation between the FFP filter and gas absorption spectra in both these applications.

We present a fibre-optic correlation spectroscopic technique, which can provide real-time, remote gas detection, identification, and radial velocity measurements using a single channel lock-in amplifier for terrestrial, atmospheric, and eventually exoplanetary remote sensing. In Section 2.1, we first present the basic operating principle in term of simulated gas absorption lines, FFP filter transmission spectra, and lock-in amplifier signal processing. In Section 2.2, we show the dependence of the lock-in signal on absorption line depth and spectral broadening. We compare the FFP filter correlation technique to stacked spectra from dispersive spectroscopy in simulation with white noise. In Section 2.3, we calculate the lock-in amplifier limiting absorption depth resolution limit. Section 3 shows the experimental demonstration of the FFP correlation technique for remote $CO_2$ detection. In Section 4, we then develop the idea of radial-velocity sensing using a lock-in amplification, and demonstrate the technique in simulation using a near-infrared (NIR) reflection spectrum of Venus.

## 2. Theory and Simulation

### 2.1 Theory

The FFP filter is a resonant cavity created by an air-gap between two cleaved single mode fibre facets, as shown in Figure 1.

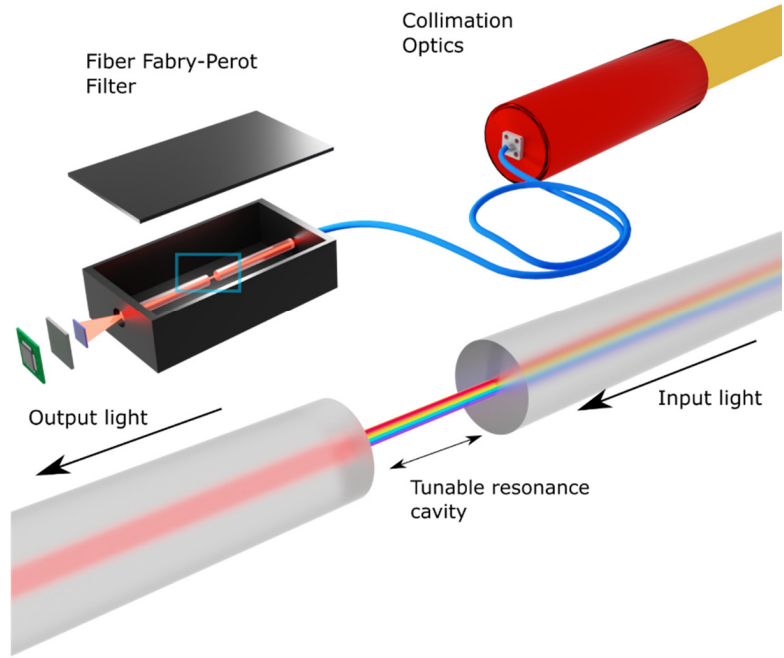

Figure 1. Illustration of the operation of a FFP filter in a free-space coupled system.

Light which passes through this cavity interferes with itself creating a resonance condition for multiples of the wavelengths of light that match the cavity length ($L$). The free spectral range ($\Delta\lambda_{FSR}$) of the fibre is directly related to the cavity length through

$$\Delta\lambda_{FSR} = \frac{\lambda^2}{2nL} \quad [1]$$

where $n$ the refractive index of the medium and $\lambda$ is the wavelength of the light. The finesse ($\mathcal{F}$) of the cavity is described by

$$\mathcal{F} = \frac{\Delta\lambda_{FSR}}{\delta\lambda} \quad [2]$$

where $\delta\lambda$ is the full width at half maximum of the resonance. The transmission spectrum of a FP mode is Lorentzian in shape [21], with a spectral profile as a function of wavelength according to

$$T_{FP}(\lambda, \lambda_0, \gamma) = \left(\frac{2}{\pi\gamma}\right)\frac{\gamma_{FP}^2}{(\gamma_{FP}/2)^2+(\lambda-\lambda_0)^2} \qquad [3]$$

where $\delta\lambda = \gamma_{FP}$ is the full width at half maximum of the transmission peak, $\lambda$ is the wavelength of light, and $\lambda_0$ is the center wavelength of the peak. For FFP filters with high free spectral ranges, adjacent resonances can be separated by >100 nm in bandwidth so that only a single transmission window is visible over a large wavelength range. For simplicity, the input light to the FFP filter is assumed to be a spectrally uniform broadband background spectrum overlaid with a gas absorption spectrum.

The simulated gas absorption spectrum of a single line takes the shape of a scaled Voigt profile subtracted from a flat background, described by

$$A(\lambda, \lambda_0, \sigma_g, \gamma_g) = 1 - \alpha_s V_g(\lambda, \lambda_0, \sigma_g, \gamma_g) \quad A:[0,1] \qquad [4]$$

where $V_g$ is a Voigt profile of the gas line, $\sigma_g$ is the Gaussian standard deviation, $\lambda_0$ is the center wavelength, $\alpha_s$ is an absorption scaling parameter, and $\gamma_g$ is the Lorentzian scale parameter. The transmission of the wavelength of the FFP filter can be dynamically tuned through the piezoelectric actuator, creating a wavelength dependent transmission function of the filter that can be temporally modulated.

The product overlap between the absorption line spectrum and the time dependent FFP transmission spectrum is expressed as

$$S_{FP}(\lambda, \lambda_0) = A(\lambda, \lambda_0, \sigma, \gamma_g) \cdot T_{FP}(\lambda, \lambda_0, \gamma_{FP}), \qquad [5]$$

where $T_{FP}$ is the transmission function of the FFP filter. The total transmitted light over a spectral bandpass from $\lambda_1$ to $\lambda_2$ with an ideal single-channel photodetector (assuming unity quantum efficiency) is the integral of the product above:

$$C(\lambda_0) = \int_{\lambda_1}^{\lambda_2} S_{FP}(\lambda, \lambda_0)\, d\lambda. \qquad [6]$$

*2.2 Simulation of $CO_2$ detection*

We choose the $CO_2$ absorption spectrum to demonstrate our technique for its quasi-periodic absorption features in the telecommunication C-band where optical fibre technologies are mature, $CO_2$ is relevant for atmospheric and exoplanet studies [33–36], and $CO_2$ relatively safe to handle. Light is coupled from free-space into a single mode optical fibre using a collimator.

The output spectrum from the filter is calculated as the product of the FFP filter transmission and the normalized gas absorption spectrum. For speed, the Voigt profile is computed for each absorption line of a gas using a rational approximation by using a rapid MATLAB subroutine [37]. The absorption spectrum of the gas is generated by importing absorption cross-sections from the high-resolution transmission molecular absorption database (HITRAN) and generating approximate Voigt profiles based on room temperature and pressure conditions. The FFP filter transmission profile is calculated as a Lorentzian function. The FFP filter transmission central wavelength is modulated around the $CO_2$ spectral features, yielding a modulated output correlation signal that is a function of the spectral feature depth, wavelength, and broadening. Simulated output spectra $S_{FP}(\lambda)$ produced from a single $CO_2$ absorption line and the FFP filter transmission for center wavelengths ranging from $\lambda_0$= 1578.575 to 1578.660 nm are shown in Figure 2. The integrals of the $S_{FP}(\lambda, \lambda_0)$ curves in Figure 2 correspond to the correlation signals $C(\lambda_0)$ for each center wavelength.

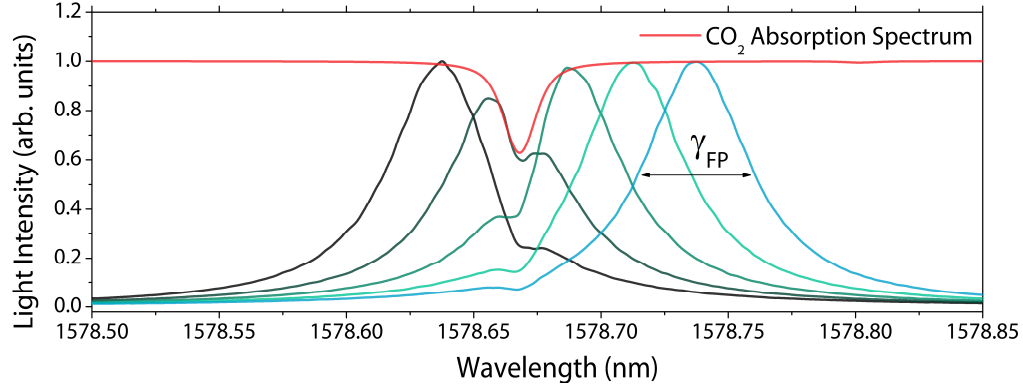

Figure 2. Simulated transmission spectrum of the tunable Fabry-Pérot filter as it is tuned across the single $CO_2$ spectral line at 1578.66 nm (red).

For the simulation of $CO_2$ detection using the absorption lines near 1580 nm, we introduce a periodic modulation of the simulated FFP filter transmission central wavelength. We assume no noise for the purposes of this simulation. The lock-in amplification is simulated according to Eq. 7 as the transmission central wavelength $\lambda_0$ of the FFP filter is scanned in time. The signal output from a lock-in detector is described as

$$V_{out} = \frac{1}{T_c} \int_{t-T_c}^{t} \sin(2\pi f_{ref} t^* + \varphi) V_{in}(t) dt^* \qquad [7]$$

where $\varphi$ is the phase shift, $t$ is the time in discrete simulation steps, $T_c$ is the lock-in time constant, and $f_{ref}$ is the reference frequency. The modulation period is $T_{mod} = \frac{1}{f_{ref}}$. The modulation was configured as a sawtooth wave to ensure the transmission filter overlaps the absorption feature only once per period even if the Doppler shift of the spectral line is unknown. A sinusoidal modulation can be used if the Doppler shift is negligible and the modulation is set such that an extremum of the sweep correspond to the maximum overlap of the resonance with the gas absorption line. The center wavelength is related to the time by the sawtooth function

$$\lambda_0(t) = 2\left(\frac{t}{T_{mod}} - \left[\frac{1}{2} + \frac{t}{T_{mod}}\right]\right). \qquad [8]$$

The input signal to the lock-in amplifier is therefore equivalent to the correlation signal,

$$V_{in}(\lambda_0) \equiv C(\lambda_0) \text{ and } V_{in}(t) \equiv C(t), \qquad [9]$$

Lock-in amplification of the input correlation signal was simulated using a discretized version of Equation 7. For the purposes of the simulation, time is discretized into steps of 20 per modulation period. The input signal $V_{in}$ was calculated at each step using Eq. 4. The output signal was calculated over a range of $\varphi$ to find the maximum lock-signal. The time constant ($T_c$) was set at 900 simulation steps. The modulation signal is shown in Figure 3 at a range of absorption line depths. The amplitude of the transmission wavelength modulation was set at 0.47 nm, which is approximately the absorption line spacing of $CO_2$ near 1579 nm. The spectral lines were generated using a rational approximation to the Voigt function [37], with parameters of $y$=0.5 and a wavelength scaling parameter of 100 to generate approximate Voigt functions with FWHM of ~30 pm. This spectral profile simulates line broadening with Lorenztian (Doppler) and Gaussian (pressure) contributions. The result is a periodic modulation signal at the reference frequency of the sawtooth modulation of the FFP filter transmission center wavelength, shown in the blue dashed line in Figure 3. The modulated correlation signal corresponds to the variation in transmission through the FFP scanning through a single $CO_2$ absorption line centered at 1579.1 nm.

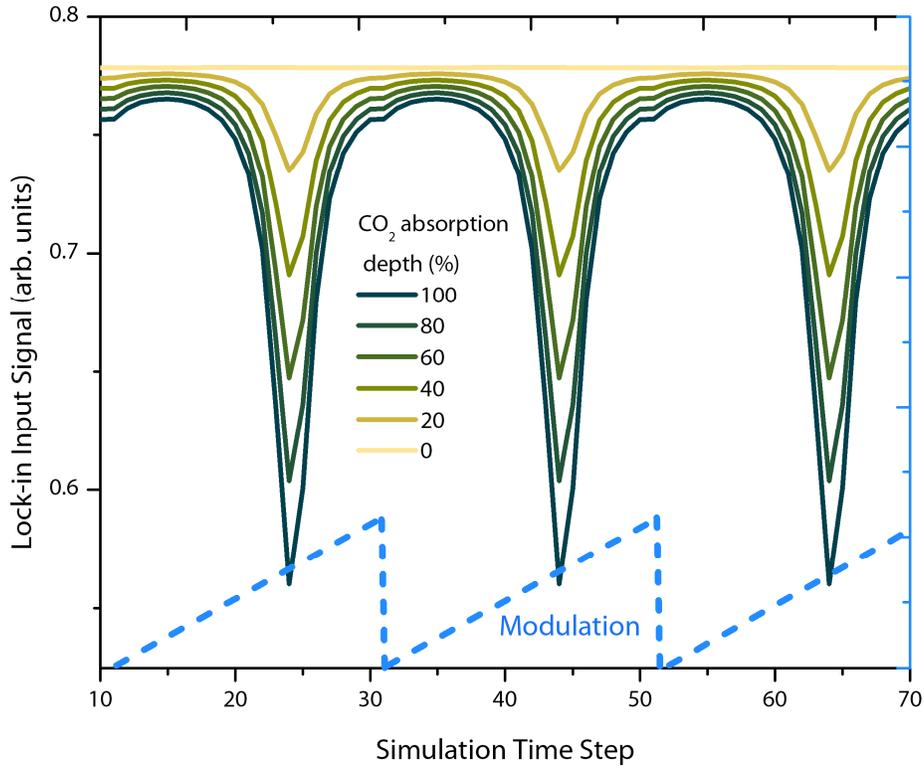

Figure 3: Simulated lock-in amplifier input signal ($V_{in}$) over time for various $CO_2$ absorption depths. The sawtooth modulation function is shown as a dashed blue line.

The minima correspond to the maximum overlap condition between the gas absorption line and the FFP filter. We calculated $V_{out}$ at a range of phase offsets and find the phase at which the signal is strongest. The phase offset is dependent on the FFP filter center wavelength, and can be used to uniquely indicate the presence of $CO_2$ and measure the Doppler shift of the $CO_2$ line. To minimize the occurrence of false positives, the center wavelength modulation should be in a spectral range where there are no gas lines from other common or expected gas species.

The correlation technique is sensitive to the gas pressure due to the broadening of the spectral line. We show the variation in the lock-in input and output signals with increasing spectral line full-width at half maximum (HWHM) by changing the wavelength scaling parameter in Figures 4 and 5, respectively. For Figure 4, the FFP transmission bandwidth is fixed at 50 pm and the $CO_2$ absorption depth is set at 50%. The lock-in input signal modulation is broadened and deepened with increasing spectral line FWHM, indicating improved detection contrast. The lock-in amplifier DC output signal, which is related to the magnitude of the modulation, is maximized at around 150 pm, and decreases at even higher temperature and pressures as the spectral lines begin to overlap. The technique is the most sensitive at higher gas pressures and temperatures where spectral lines are still well-separated and the overlap between the FFP filter transmission windows and the spectral line is maximized.

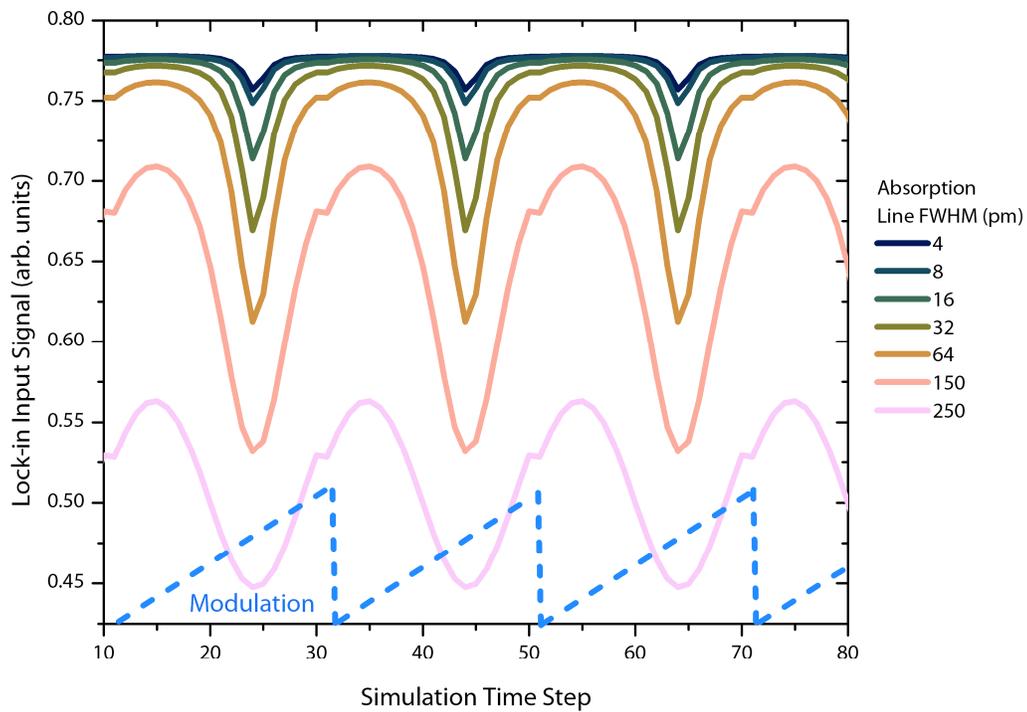

Figure 4: Simulated lock-in amplifier input signal ($V_{in}$) over time for various $CO_2$ absorption line FWHM values. The modulation signal is shown as a blue dashed line. The additional periodic features at time steps 31, 51, and 71 are artifacts of the sawtooth modulation.

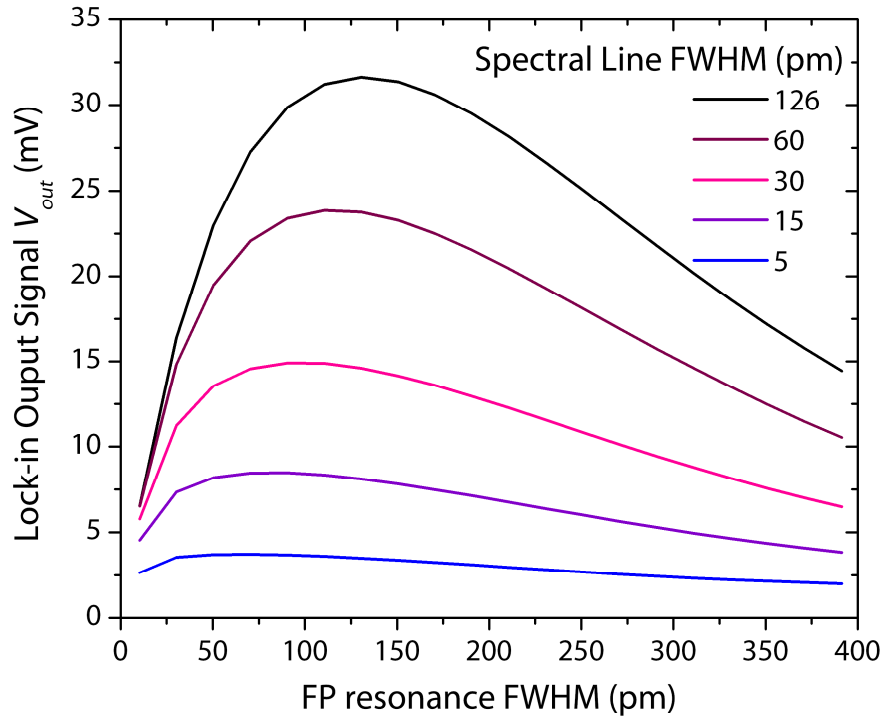

Figure 5: Simulated lock-in amplifier output ($V_{out}$) signal as a function of the absorption line FWHM. The voltage of the lock-in amplifier is arbitrary in scale as it depends on the signal gain.

The sensitivity of the lock-in amplifier correlation technique depends on the time constant, the absorption depth of gas lines, noise, and the modulation characteristics of the filter. We simulated the sensor response with increasing $CO_2$ absorption depth, and show the correlation signal in Figure 6a. The noise was set at 5% of the broadband signal, with an absorption depth ranging from 0 to 5% to reproduce low SNR values (up to SNR=1). The time constant is set at 4900 steps with a total of 5000 time steps. The signal from the last 100 steps were averaged to produce the final $V_{out}$ value. The output voltage from the lock-in amplifier increases with absorption depth with a slope of $1.75 \times 10^{-5}$ V/% absorption depth.

We compare the signal from the simulated lock-in with a simulated stacked spectrum (Figure 6b) obtained from dispersive spectroscopy of the FPP filter throughput but otherwise with the same noise characteristics. The sensitivity of the two techniques are comparable with detection of the signal being detectable at ~0.5%, despite the FFP filter only accepting photons from a narrow bandpass. The main advantages of the FFP filter-based method are that it is less sensitive to cosmic ray events, can be significantly more compact and lower cost, produces much less data (thus requiring less data processing), and can track a gas signal in real-time.

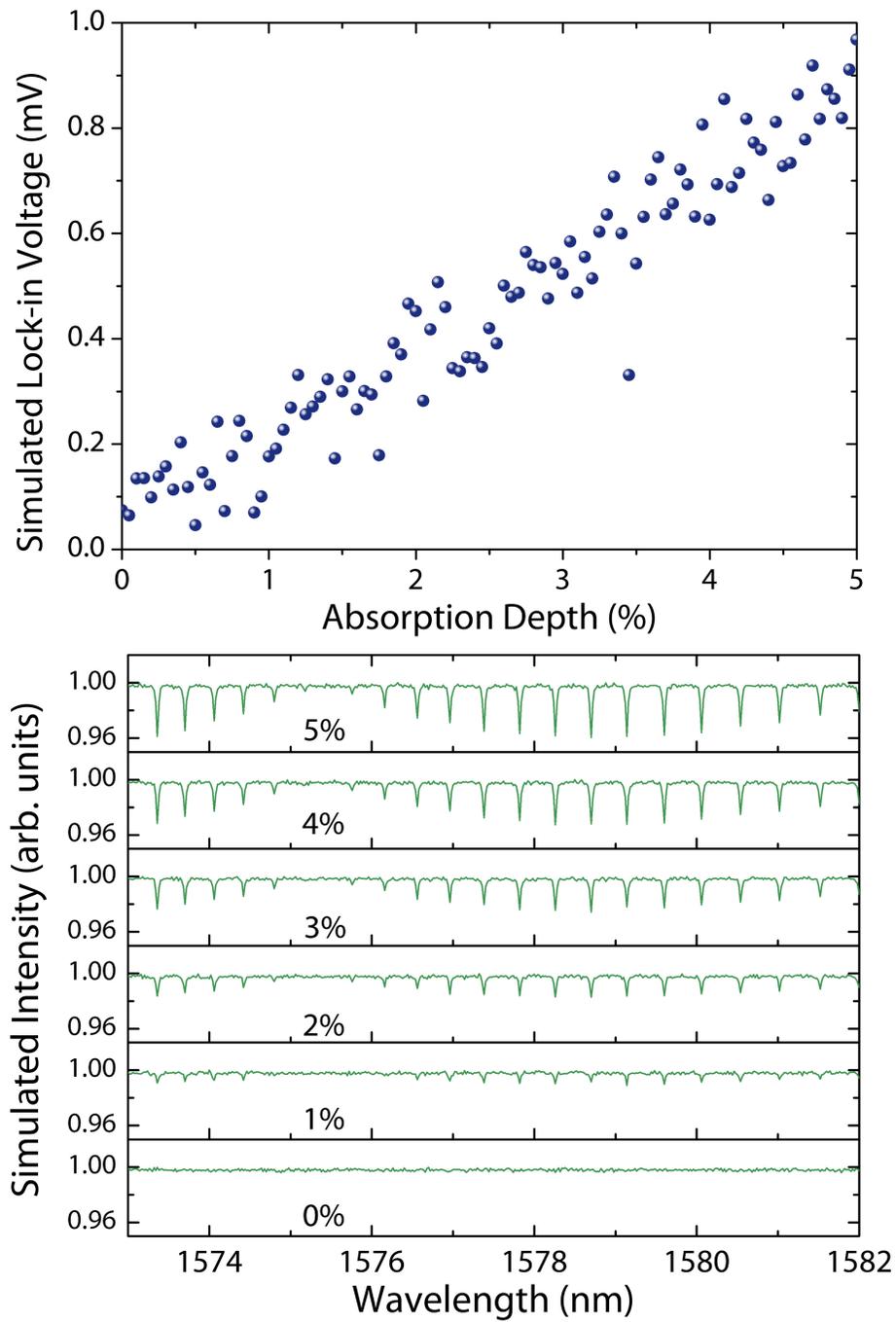

Figure 6: (a) Simulated lock-in amplifier output ($V_{out}$) signal as a function of the absorption depth of the $CO_2$ line. (b) Simulated noisy spectra stacked over the same number of time steps for a range of absorption depths.

*2.3 Detection limit*

For these parameters used in Section 2.2, detection limit for changes in the absorption depth is on the order of ~0.5% due to the added white noise. However, the lock-in amplifier input noise and detector noise also provide a limit to the absorption depth resolution. With sufficient light, the fundamental limit to the sensitivity is expressed as

$$SNR = 1 = \frac{AM^V_{RMS,min}}{(NEV + N_{lock})\sqrt{\Delta f}} \quad [10]$$

where $AM^V_{RMS,min}$ is the lock-in signal amplitude for the minimum gas absorption depth detectable. For our Zurich Instruments MLFI lock-in amplifier, we have input voltage of $N_{lock}$ = 40 nV/√Hz at 1 Hz, and our InGaAs avalanche photodiode has a noise equivalent voltage (*NEV*) of 100 nV/√ Hz at 1 Hz. The noise bandwidth is approximated as $\Delta f \approx 1/(4\tau_c)$ to be below the Nyquist frequency, preventing aliasing of the modulation signal. We assume a time constant of 10 seconds, representing a reasonably long time interval where systematic drift could be experimentally managed and minimized. The simulation results provided a sensor sensitivity of $A^{V/\%}_{sens}$ =1.75 ×10$^{-5}$ V per percent absorption depth ($A_d$) of the room temperature and pressure broadened gas absorption line. Using

$$A^{V/\%}_{sens} \times A^{\%}_{d,min} = AM^V_{RMS,min} \quad [11]$$

we can solve for $A^{\%}_{d,min}$ (the absorption depth resolution) using Equation 10 and Equation 11. We find an approximate minimum detectable absorption depth of ~10$^{-4}$ % using our current lock-in amplifier and detector. This corresponds to a resolution of a few ppm over a gas column length of 10 m. This value can be further improved with cryogenics, lower-noise electronics, and longer lock-in amplifier time constants. Time constants are limited to hundreds of seconds or less in practice, and may be further limited by systematic drifts during observations. For most astronomical observations, low dynamic response is generally an acceptable trade-off to gain maximum sensitivity.

## 3. Experimental Detection of $CO_2$

To experimentally demonstrate the gas sensor, we constructed a fibre-coupled system which guides light through a gas, into the FFP and onto a detector. We use a fibre-coupled 1555 nm superluminescent LED as a broadband light source and a Thorlabs multi-pass Herriott cell with a 10.5 m effective path length as the gas cell. $CO_2$ was injected into the chamber to 1 atm, with the pressure being monitored by a vacuum gauge. The Herriott cell was evacuated multiple times using a roughing pump prior to re-filling with pure $CO_2$. The lock-in amplifier time constant was set at 100 ms and the lock-in frequency was set to 1 kHz. The control circuitry of the FFP filter can only generate a triangular wave pattern for the FFP central wavelength modulation, which introduces two peaks in the correlation signal, one when increasing the center wavelength, and another when decreasing. This effect is mitigated by setting the maximum center wavelength of the FFP filter modulation range to a $CO_2$ line, such that the two correlation peaks are degenerate. A function generator was used to provide the clock signal to the FFP filter driver circuit and the external reference of the lock-in amplifier. A Thorlabs APD410C InGaAs avalanche photodiode was used to detect the transmitted light. The bandwidth of the FFP filter was 220 pm (15 GHz) with a finesse of 1000.

Sufficiently low (~1 kHz or lower) modulation frequencies are required in practice to minimize the effects of electrical capacitance and mechanical inertia of the piezoelectric drive. In our experiment, a modulation frequency of 1 kHz was used. A schematic of the experimental system is shown in Figure 7.

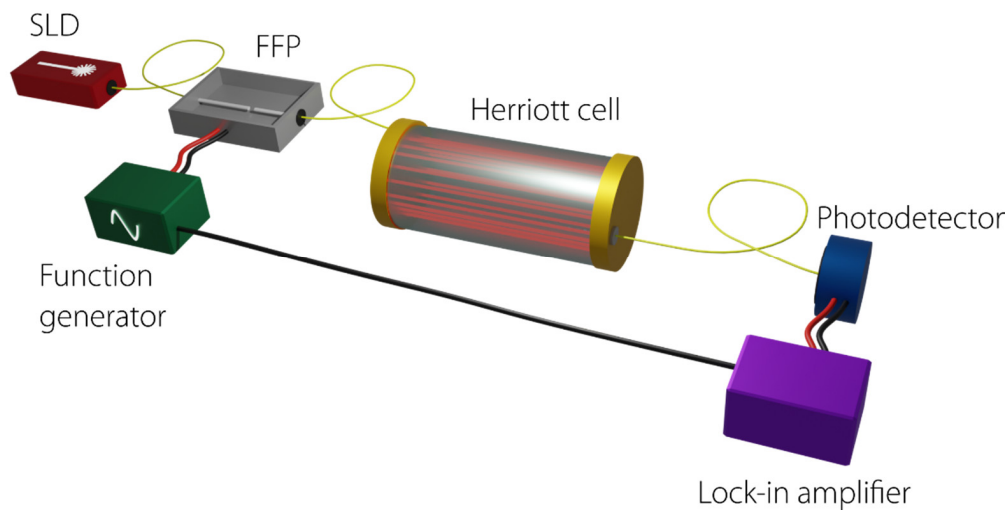

Figure 7: Schematic illustration of the optical and electrical experimental benchtop system. The grey box indicates the fiber Fabry-Pérot filter. The yellow wires represent polarization maintaining optical fibres.

We show the lock-in amplifier output voltage as a function of $CO_2$ gas cell pressure in Figure 8. We observe an approximately linear relation between pressure and $V_{out}$ output signal at lower pressures. As the gas pressure increases in the low pressure regime, the higher optical depth of the $CO_2$ absorption lines leads to a higher lock-in amplifier voltage. At medium to higher pressures, the absorption lines are mainly broadening rather than increasing in absorption depth, leading to a slightly reduced sensitivity of the sensor at higher gas pressures. The $V_{out}$ signal is calibrated in post-processing to zero at zero pressure. We observe a saturation behavior that arises because the majority of the lock-in signal increase is a result of the absorption line broadening with increased pressure. This effect has previously been observed in bulk optical FP correlation spectroscopy [14].

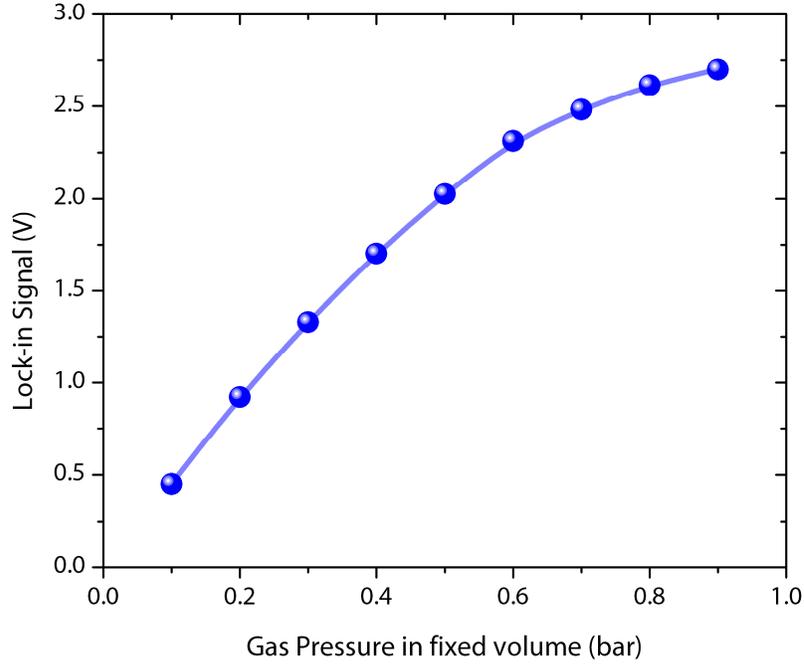

Figure 8: Experimental $V_{out}$ signal as a function of gas pressure in the Herriott cell. Pressure and lock-in signal uncertainties for the pressure and lock-in voltage are 10 mbar and 0.01 V, respectively, and are too small to be shown here. The curve is used as a guide to the eye.

## 4. Doppler shift sensing

### 4.1 Phase detection for Doppler shift estimation

One of the benefits of lock-in amplification is the ability to measure the input signal phase of a modulated input signal in additional to the signal magnitude. In the case of a sawtooth modulation pattern, the phase of the lock-in signal is proportional to the wavelength of the gas absorption line relative to the wavelength offset of the sawtooth modulation. This effect can be applied in astronomy for measuring radial velocities of celestial bodies. The two orthogonal components of the lock-in signal $V_{sig}$ can be used to obtain the phase of the lock-in signal using the equations below:

$$V_{sig} = \frac{2V_{out}}{\cos\theta} \qquad [12]$$

$$X = V_{sig} \cos\theta \qquad [13]$$

$$Y = V_{sig} \sin\theta \qquad [14]$$

$$\theta = \arctan\left(\frac{Y}{X}\right). \qquad [15]$$

In the case of known molecular absorption lines, the spectral shift is proportional to the radial velocity of the object. At detection wavelengths around 1579 nm where many closely spaced carbon dioxide absorption lines are present, the unique measurement of a Doppler shifts is limited to the spectral spacing between spectral lines (~0.4 nm to ~0.5 nm). This limits the radial velocity measurement range to ±40 km/s. Adjacent absorption lines can enter the modulation range and cause phase wrapping of the radial velocity value. We show by simulation the phase shift of the AC signal from the FFP filter with increasing radial velocity of the $CO_2$ target in Figure 9a. The FFP center wavelength was modulated using a sawtooth modulation around the $CO_2$ spectral line at 1578.7 nm for 3 periods. The phase of the signal can be determined by finding the phase at which the maximum lock-in signal occurs, using Equation 15. In Figure 9b, we show by simulation the radial velocity as a function of the phase. For a modulation range of 0.8 nm, the sensitivity of the radial velocity sensor is -212.5 (m/s)/deg and can be directly used to determine from the radial velocity of a remote target with high resolution after calibration. If the SNR is sufficiently high in a stable system, the resolution is fundamentally limited by the phase resolution of the lock-in amplifier, which is $10^{-6}$ deg for our lock-in amplifier. The resulting radial velocity resolution limit is 2 mm/s, a value that would surpass the state-of-the-art Doppler search spectrographs if systematic effects are neglected. In practice, however, the resolution may be limited by system stability, lock-in input noise, telluric contamination, and shot noise. The sensitivity can be improved by using a larger modulation range at the expense of throughput, since for a large modulation range the FFP filter transmission wavelength would not overlap the spectral line for most of the modulation period. The maximum modulation range is limited by the presence of the strong adjacent absorption lines, such as in the case of $CO_2$.

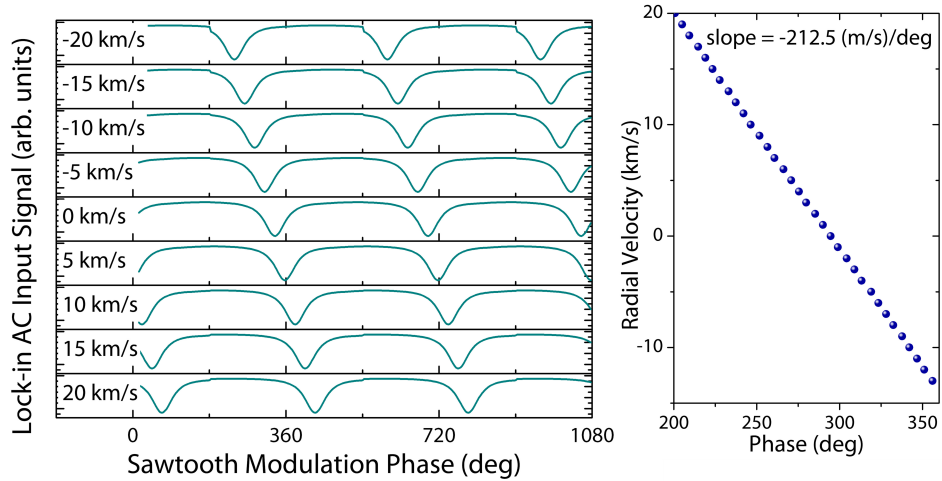

Figure 9: (a) Total light output from the FFP filter (AC input to the lock-in amplifier) as a function of the phase of the sawtooth modulation of the central transmission wavelength for a range of radial velocities. The artifact halfway through the modulation arises from the discontinuity of the sawtooth function. (b) The phase of the signal as a function of radial velocity. The slope can be used to determine the radial velocity of the target using the measured phase of the AC signal.

*4.2 Experimental reflection spectrum of Venus*

Venus is a bright planet in the night sky that can provide a $CO_2$-rich reflection spectrum for which to demonstrate FFP correlation sensing and radial velocity measurement technique in simulation with real spectral data. The radial velocity of Venus is well-known as a function of time and location on Earth. The NIR reflection spectrum of Venus can be correlated with the FFP filter transmission spectrum in simulation to demonstrate how the radial velocity can be extracted from the phase of the lock-in signal. A NIR spectrum of Venus was acquired using the 3.58-meter Galileo National Telescope on the GIANO-B echelle spectrograph [38] at La Palma and is shown in Figure 10. The spectrum was acquired on May 4$^{th}$ 2020 at 20:50:16 UTC. We used the order-reduced, order-merged spectrum from the spectrograph. The spectrograph has a resolving power of ~50,000 that corresponds to a resolution of ~32 pm at 1580 nm. The air mass (AM) was 1.76 at the time of measurement. The broadened spectral lines are an artifact of the limited spectrometer resolution. The spectrum reveals strong $CO_2$ absorption bands centered at 1580 nm and 1610 nm. A more detailed view of the absorption line profiles shows that the $CO_2$ absorption peaks of both Earth and Venus are partially overlapping, where the Venusian absorption lines are blueshifted by 60 pm. The telluric lines are corrected by dividing by an A-class star telluric absorption reference spectrum of Earth with AM = 2.

As an example, the FFP filter center wavelength can also be rapidly modulated with a small amplitude and gradually scanned in wavelength over the absorption features. With a narrow bandwidth FFP filter, Earth's telluric absorption lines can be directly distinguished from that of the target since this method produces a full spectrum as in dispersive spectroscopy. The drawbacks of this technique are slower acquisition times and reduced Doppler shift resolution depending on the scan rate. Alternatively, changes in the radial velocity can be detected by looking for deviations of the phase from a nominal reference value that has only telluric absorption lines. Since the Doppler shift from Earth's $CO_2$ absorption lines is negligible during an observation, the effect of the target's $CO_2$ lines can still be detected but is more difficult to qualify since the extracted phase would be related to both Earth's and the target's $CO_2$ signals. These complexities can be mitigated by spaced-based detection or targeting of a molecule that is not a trace gas of Earth's atmosphere.

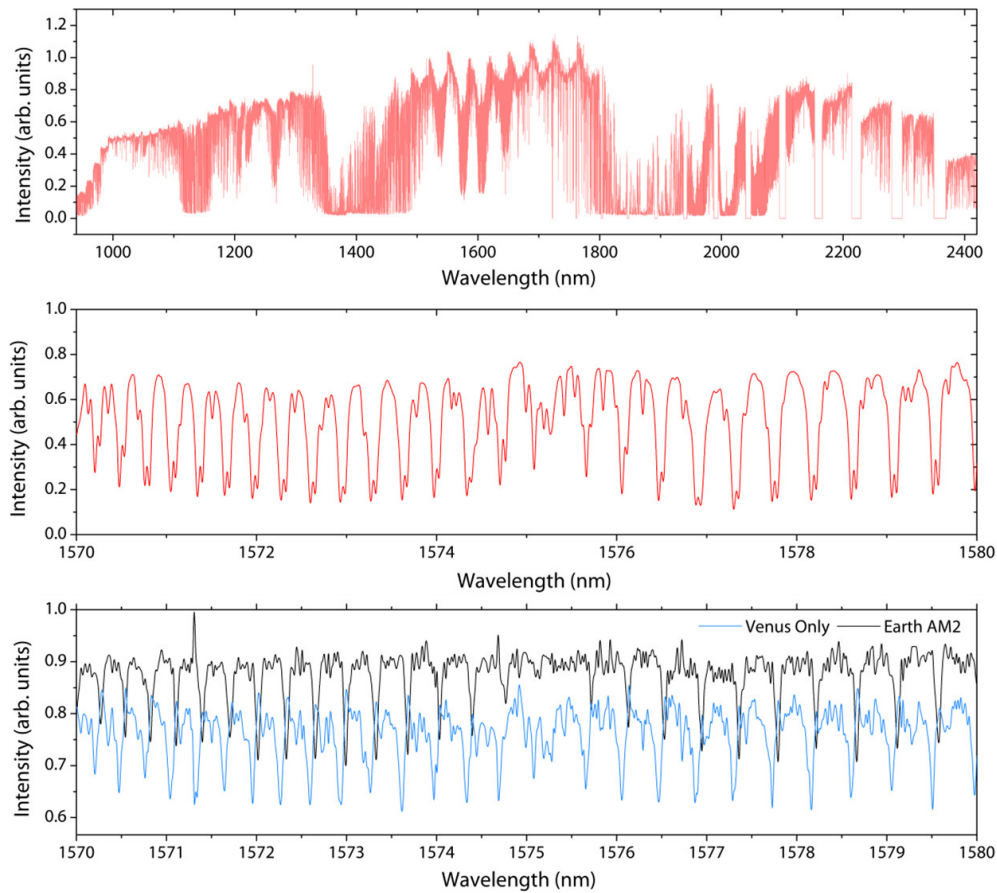

Figure 10: (a) Near-infrared spectrum of Venus taken at an air mass of ~2 through Earth's atmosphere. (b) Measured high-resolution spectrum including contribution from both Earth and Venus. (c) Processed spectrum of Venus (blue) after correcting for the telluric lines using a measured AM2 spectrum.

In the next section, we use the processed Venusian reflection spectrum as an input spectrum into the FFP filter to demonstrate correlation-based sensing for $CO_2$ in a complex absorption spectrum. For simplicity, we have used the processed spectrum where the telluric lines have been removed to simulate space-based observations.

### 4.3 Higher order harmonic radial velocity sensing

In the case of an ideal, single gas absorption spectrum, the radial velocity can be accurately determined by the phase shift of the lock-in signal. However, the radial velocity to phase relationship is affected by other spectral lines in the modulation range (such as with the Venusian reflection spectrum), leading to deviations in the radial velocity determination unless a calibration for that same spectrum is performed. Since the spectrum in most cases is not known, this can prevent the accurate determination of the radial velocity from complex spectra.

For example, a CO line near a $CO_2$ line can significantly affect the detection and phase measurement.

As there are multiple, quasi-periodically spaced absorption lines for many gases such as $CO_2$ and CO, these additional lines can be used to increase the selectivity of the gas sensing technique and improve the accuracy of the radial velocity measurement. By scanning a range of a multiple of the average line spacing ($\Delta\lambda_{mod} = n_h \times \Delta\lambda_f$), the higher order harmonic frequencies of the lock-in of $f_h = n_h \times f_r$, where $f_r$ is the reference lock-in frequency and $n_h$ is the harmonic order, would also indicate the presence of the gas provided the harmonic order matches the number of nearly equally spaced spectral lines in the scanning region. Figure 11 shows the possible modulation ranges and corresponding lock-in harmonic order overlaid on the $CO_2$ absorption spectrum. This technique increases the specificity of the detection and is less sensitive to spectral features from other gases since their contribution is reduced relative to the signal from the target gas. This technique produces nearly the same sensitivity as the single line correlation technique if the single line modulation range is the same as the spacing between absorption lines. The unequal spacing between adjacent gas lines limits the modulation range of the technique to a maximum of 4-8 absorption lines in the case of $CO_2$, depending on broadening of the spectral lines and the FFP filter FWHM.

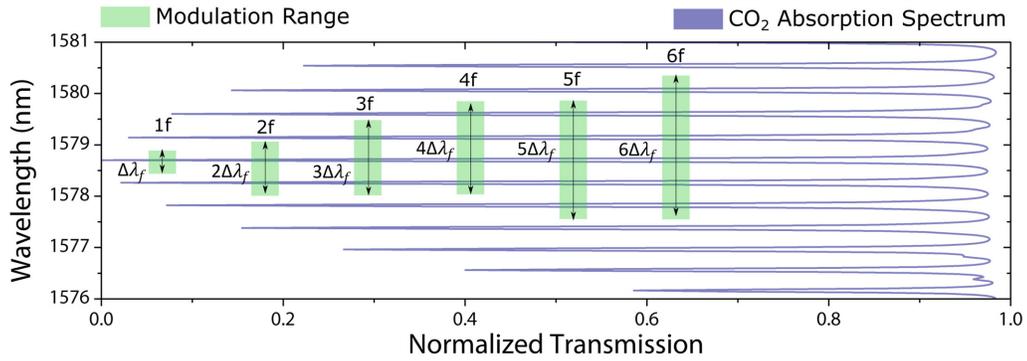

Figure 11: (a) Absorption spectrum of $CO_2$ overlaid with FFP filter modulation ranges for higher order harmonic radial velocity sensing. The height of the green bars represent multiples of the fundamental modulation range, with the harmonic frequency marked above each bar.

We simulate the extraction of the sixth harmonic signal $6\times f_r$ from a modulation range that is six times the average spacing between $CO_2$ lines ($\Delta\lambda_{mod} = 6 \times 0.46$ nm $= 2.76$ nm), and show the phase as a function of radial velocity in Figure 12. Different gas absorption models including the Venusian spectrum were introduced to demonstrate the sensitivity to molecular species. The $CH_4$ correlation was simulated at a wavelength region around 1650 nm. The different gases are distinguished by their phase offsets relative to a calibrated phase. The phase shifts for $CH_4$ and CO are distinct from that of $CO_2$. The phase offset of the $CH_4$ signal can be guaranteed through the appropriate choice of the center wavelength of the sawtooth modulation. When more than one gas species is present, as we show for a 50% mixture of CO and $CO_2$, the lock-in output phase is an interpolation of the two phases from each of the single gas spectra, with the measured phase being a function of the contribution of each gas to the total absorption over the modulation wavelength range of the FFP filter. This makes it possible to use the phase offset to infer relative gas ratios if the radial velocity is known. For the pure $CO_2$ spectrum, the spacing between the strongest lines is approximately $\Delta\lambda_f = 0.46$ nm. We simulate the 6$^{th}$ harmonic order correlation with $\Delta\lambda_f = 0.43$ nm, 0.45 nm, and 0.47 nm to assess the sensitivity of the sensor to the fundamental $CO_2$ wavelength separation. The phases are very similar at slightly different values of $\Delta\lambda_f$. Radial velocities should be calibrated for a specific value of $\Delta\lambda_f$ since the phase is directly related to the modulation range $\Delta\lambda_{mod}$. The spectrum of Venus, which contains numerous other spectral features, shows nearly identical phase as the pure $CO_2$ signal since the spectrum is dominated by the $CO_2$ absorption lines. The capturing of the 6$^{th}$ order harmonic signal is able to average the contributions of other spectral features that occur at non-matching wavelength spacings to $CO_2$ such that they are suppressed relative to the $CO_2$ contributions.

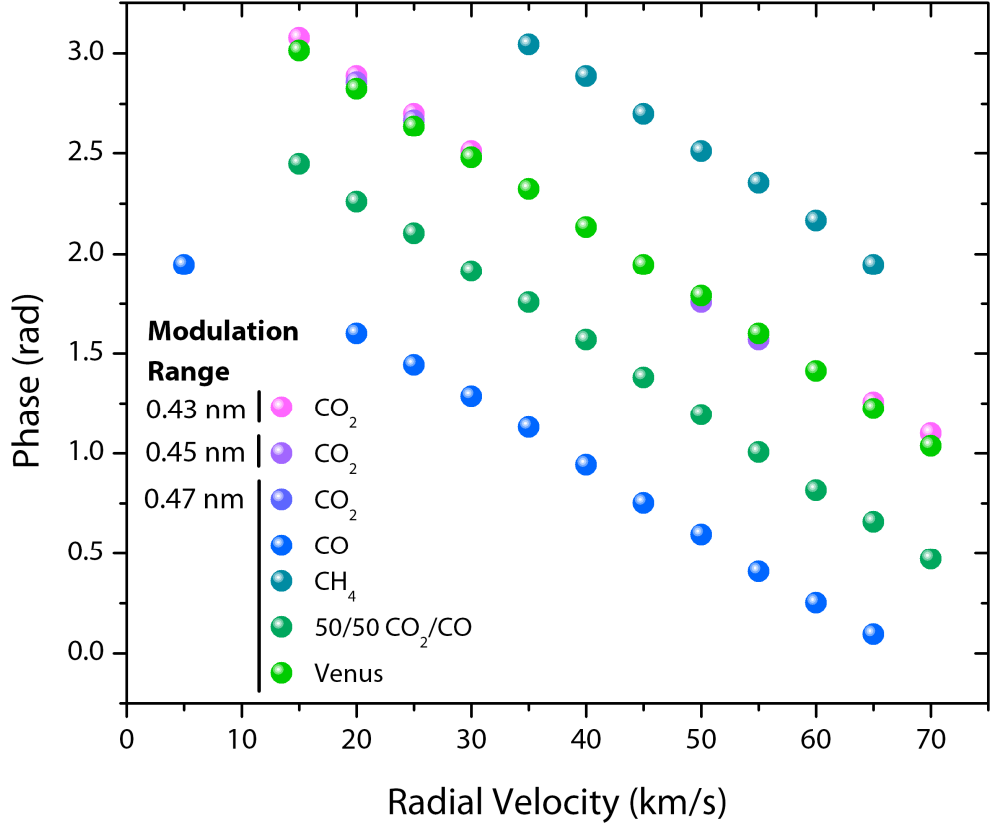

Figure 12: Phase shift as a function of radial velocity of various redshifted spectra using the 6$^{th}$ order lock-in amplifier harmonic over a modulation range of 6×$\Delta\lambda_f$ and a harmonic order of 6 times the average spacing between adjacent $CO_2$ lines (2.83 nm) centered at 1580.2 nm. The value of $\Delta\lambda_f$ is varied from 0.43 nm to 0.47 nm for the $CO_2$ spectrum.

We demonstrate this radial velocity and gas species detection technique by numerical simulation using the observed Venus spectrum. Calibration of the sensor with a reference $CO_2$ spectral lines is required to provide a reference phase signal for radial velocity determination. A sawtooth modulation function is used. A phase shift relative to the reference phase can be associated to a Doppler shifted feature, whereas a phase shift using $n_h = 1$ could be a result of another spectral line. In Figure 12, the phase of the lock-in signal is plotted as a function of simulated radial velocity for various spectra. We show that for various $CO_2$-tailored modulation ranges, the phase shift as a function of radial velocity is relatively unchanged. The reflection spectrum of Venus is nearly indistinguishable to that of pure $CO_2$, due to the strong $CO_2$ absorption lines in the spectrum, showing that the technique can operate under complex spectra such as that of Venus.

It should be noted that the lines in the observed spectrum of Venus are broadened due to the instrumental resolution of GIANO-B (R~50,000). This decreases the radial velocity accuracy. The radial velocity of Venus at the time and place of spectral acquisition by the TNG telescope was determined to be -10.92 km/s using the NASA JPL Solar system dynamics HORIZONS web interface [39]. We overlap the measured NIR spectrum of Venus with the transmission window of the FFP filter to simulate the output lock-in amplifier signal. The phase offset is extracted and compared to the phase offsets using a pure $CO_2$ spectrum over a range of radial velocities. The radial velocity at which the $CO_2$ phase offset matches that of Venus spectrum indicates the approximate radial velocity of Venus. The radial velocities were determined with modulation ranges from 1-6 line spacings and shown in Figure 13. Our radial velocity estimation accuracy for Venus did not improve with additional absorption lines in the modulation range. Possible reasons for this could be due to errors in the telluric correction or additional absorption lines in the Venusian reflection spectrum that leads to deviations from the accepted value. A more accurate strategy was employed to improve accuracy, where the signals from all harmonic multipliers were averaged to produce a radial velocity estimate of approximately -10.52 km/s, which has a relative error of 3.7% from the true value obtained using the accepted value of -10.92 km/s from [39]. Errors from the radial velocity estimation at each harmonic multiplier arise from other absorption features in the spectral range, so the expected uncertainty is dependent on the strength of the $CO_2$ relative to the other spectral lines. While obtaining a precise value of the radial velocity may be challenging, observing changes in the radial velocity are less dependent on the contribution from the telluric lines and could, for example, be used for exoplanet discovery or confirmation.

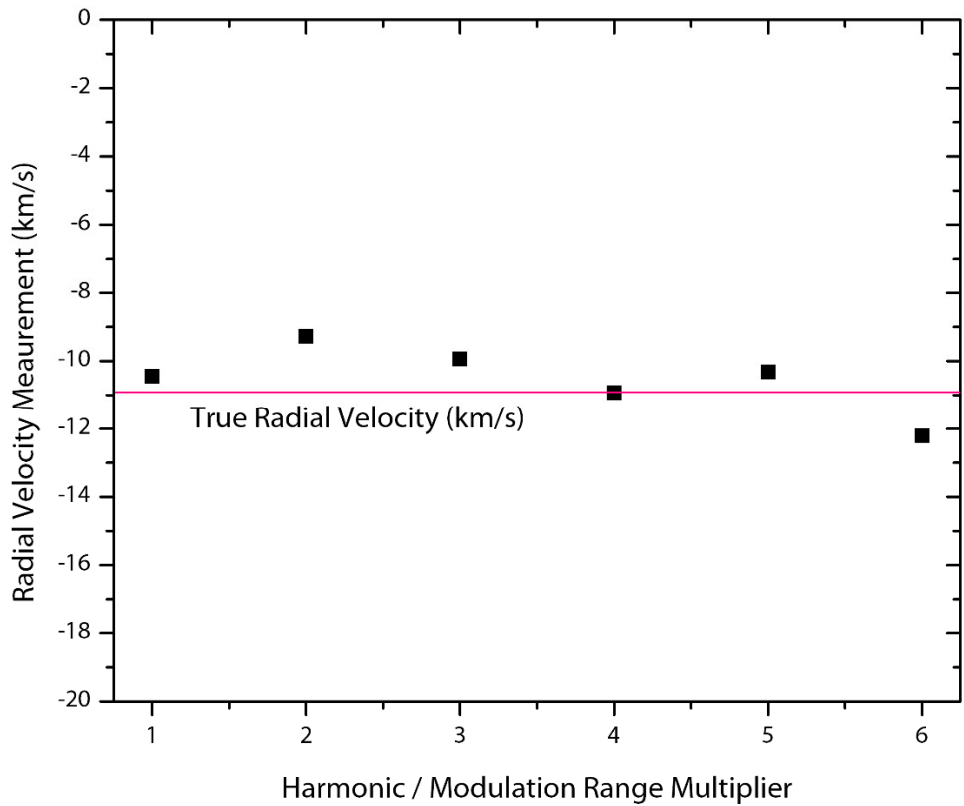

Figure 13. Radial velocity extracted from the Venus spectrum for increasing lock-in demodulator harmonics $n_h$. The true radial velocity of Venus at the time of acquisition is shown as the colored line.

## 5. Discussion

Spectral correlation sensing using a FFP filter has numerous advantages when compared to dispersive spectroscopy in terms of detector noise, throughput, cost, size, data simplicity, and ease of high-resolution radial velocity estimation. The throughput of the FFP is >50% at the center wavelength, which is much higher than a typical spectrometer with diffraction gratings, mirrors, lenses, and additional optics. Fibre-based components are compact and low-cost and can be readily sourced commercially. High frequency noise is suppressed by the lock-in amplifier, resulting in stable, sensitive measurements when using long lock-in time constants ($T_c$). Furthermore, the single element photodetector reduces data complexity and noise compared to using detection arrays.

On the other hand, the radial velocity detection scheme has certain limitations when compared to dispersive spectroscopic techniques. First, the dynamic range of the radial velocity measurement is limited by the spacing between adjacent absorption lines, which depends on the gas species. If the modulation range is increased to include a second absorption line using the fundamental lock-in demodulation frequency, the phase shift measurement is related to both spectral lines and is not proportional to the radial velocity.

Second, the accuracy of the radial velocity estimation is limited by the complexity of the absorption spectrum. With dispersive spectroscopy, however, numerical correlation can be made from the complete spectrum. With FFP filter-based correlation spectroscopy, only a few spectral features are being correlated. The multiple harmonic detection technique we present can improve radial velocity accuracy with complex input spectral by correlating with additional absorption lines in the bandpass.

Furthermore, while the multiple harmonic detection technique can lead to improved gas species selectivity, throughput can suffer. When the FFP transmission window is not overlapping a gas absorption line during a multiple absorption line modulation period (which would typically be for the majority of the acquisition time) no spectral information is present and results in a lower lock-in amplifier output SNR. The use of the multiple harmonic technique should be used only when the radial velocity is either large, unknown, or the target is bright. Otherwise, a smaller modulation amplitude around the spectral feature can lead to improved instrument sensitivity to the expense of phase resolution and radial velocity measurement range. Multiple absorption line scanning should be used to further confirm gas detection since the detection signal will persist at all harmonic multipliers.

In summary, FFP filter-based correlation spectroscopy enables real-time detection and radial velocity estimation at a higher sensitivity than dispersive spectroscopy, while benefitting from lower instrument cost and size. We demonstrate the concept, simulation and experimental demonstration of a simple FFP filter correlation spectroscopy technique which can detect remote gases through their absorption features in the NIR. We show through simulated phase detection of lock-in amplifier signals that such a technique can identify gases in a $CO_2$ spectrum, CO spectrum, a mixture of CO and $CO_2$, $CH_4$ as well as with a measured near-infrared spectrum of Venus through the nominal detected phase difference. We confirm the radial velocity of Venus, which is predominantly composed of $CO_2$, through the phase shift of the lock-in signal. This technique leverages the low-noise advantages of lock-in amplification and single channel detector to deliver a sensor tailored to $CO_2$ detection. The multiple harmonic technique can be adapted to suit many other gases with quasi-periodic absorption lines by adjusting modulation amplitude of the FFP and the demodulation order of the lock-in amplifier.

Such a technique can be extended to the visible or mid-infrared to support the detection of other gases or increased sensitivity. Additional sensitivity can be obtained by also processing the reflected light from the FFP filter in the same method as the transmitted light. For star surveys, exoplanet detection and atmosphere characterization, the slow response time when using long time constants in lock-in amplification is not a concern. The limited acceptance angle of large single-mode fibre-coupled telescopes is suited for characterization of narrow field targets such as single stars and their exoplanets using FFP filter correlation spectroscopy. Higher remote gas sensing sensitivity can lead to lower cost stellar classification in surveys, and the discovery of the spectral signatures in Earth-like exoplanet atmospheres. Other possible applications for this sensing technique are in remote sensing for industrial emissions and agricultural monitoring.

## 6. Disclosures

The authors declare no conflicts of interest.

## 7. Data Availability

Data from this work may be obtained from the authors upon reasonable request.